\documentclass{jfm}
\usepackage{graphicx,color}
\usepackage{epstopdf, epsfig}
\usepackage{bm}
\usepackage{subcaption}
\usepackage{amsmath}
\usepackage[colorlinks]{hyperref}
\providecommand*{\rmd}{\mathrm{d}}

\shorttitle{Three-dimensional streaming in a Hele-Shaw cell}
\shortauthor{X. Zhang and B. Rallabandi}

\title{Three-dimensional streaming around an obstacle in a Hele-Shaw cell}

\author{Xirui Zhang\aff{1} and 
 Bhargav Rallabandi\aff{1} \corresp{\email{bhargav@engr.ucr.edu}}}

\affiliation{\aff{1}Department of Mechanical Engineering, University of California,
Riverside, CA 92521, USA}

\begin{document}

\maketitle

\begin{abstract}
The application of oscillatory flow around an obstacle drives a steady “streaming” due to inertial rectification, which has been used in a host of microfluidic applications. While theory has focused largely on two-dimensional (2D) flows, streaming in many practical microfluidic devices is three-dimensional (3D) due to confinement. We develop a three-dimensional streaming theory around an obstacle in a microchannel with a Hele-Shaw like geometry, where one dimension (depth) is much shorter than the other two dimensions. Utilizing inertial lubrication theory, we demonstrate that the time-averaged streaming flow has a three-dimensional structure. Notably, the flow changes direction across the depth of the channel, which is a feature not observed in less confined streaming setups. This feature is confirmed by our experiments of streaming around a cylinder sandwiched in a microchannel. Our theory also predicts that the streaming velocity decays as the inverse cube of the distance from the cylinder, faster than that expected from previous two-dimensional approaches. We verify this decay rate quantitatively using particle tracking measurements from experiments of streaming around cylinders with different aspect ratios at different driving frequencies.
\end{abstract}

\begin{keywords}
Authors should not enter keywords on the manuscript, as these must be chosen by the author during the online submission process and will then be added during the typesetting process (see http://journals.cambridge.org/data/\linebreak[3]relatedlink/jfm-\linebreak[3]keywords.pdf for the full list)
\end{keywords}

\section{Introduction}

Streaming refers to a secondary steady flow driven by the inertia of primary oscillations  of a viscous fluid \citep{boluriaan2003acoustic,wu2018acoustic}. The generation of streaming requires a spatially varying oscillatory flow, which may be established by the propagation of acoustic waves, by oscillations of an interface, or by bulk oscillation of fluid around an obstacle \citep{eckart1948vortices,bolanos2017streaming,karlsen2018acoustic}. Streaming flows around obstacles in micro-scale confinement typically have vortical structures and have been widely used in microfluidic applications due to their fast flow speeds, simplicity of generation, and versatility. For instance, eddies of acoustic microstreaming can be used to modify the composition of polydisperse suspensions by trapping smaller microparticles and releasing particles of larger size \citep{stone2001microfluidics,beebe2002physics,lutz2003microfluidics,thameem2017fast} or to control vesicle deformation and lysis for bioengineering applications \citep{marmottant2003controlled,marmottant2008deformation,tandiono2012sonolysis}. Acoustic streaming induced by a surface acoustic waves (SAWs) has been widely utilized to enhance mixing in laminar flow in microchannels \citep{westerhausen2016controllable,ahmed2019surface}, precisely control the intercellular distance and spatial arrangement of cells \citep{guo2015controlling,mutafopulos2019traveling}, and in the design of ``acoustic tweezers'' that can trap suspended microparticles. \citep{shi2009acoustic,ding2012chip,zhu2021acoustohydrodynamic}.

Streaming around cylindrical obstacles (cylinder or bubble) is well understood under a two-dimensional framework \citep{holtsmark1954boundary,stuart1966double,riley1967oscillatory,bertelsen1973nonlinear,hamilton2003acoustic,chong2013inertial,vcervenka2016variety,lei2018effects} and has been verified experimentally \citep{holtsmark1954boundary,bertelsen1973nonlinear,andrade1931circulations,lutz2005microscopic}. The flow is organized into four symmetric pairs of vortices around a rigid cylinder; each vortex pair consists of an inner and an outer vortex whose relative size depends on the dimensionless Stokes layer thickness  \citep{holtsmark1954boundary,lutz2005microscopic}. \citet{vishwanathan2019steady} recently reported a method to generate and control the size of these streaming vortices in a microchannel under sub-kilohertz oscillation frequencies with a loudspeaker. 

Most modern experimental realizations of streaming, for example in microfluidics, occur under vertical confinement by walls, leading to three-dimensional flows. Using experiments of streaming around an axially confined cylinder, \citet{lutz2005microscopic} reported three-dimensional streaming flows near the two ends of the cylinder and close to the cylinder surface, which converged to quasi-two-dimensional flow far away from the cylinder surface but close to the axially central plane. \citet{marin2015three} conducted experiments showing that semi-cylindrical bubble oscillations under axial confinement in a microfluidic channel also produce three-dimensional streaming flows. Measuring the trajectories of tracers using astigmatism particle-tracking velocimetry (APTV), they reported a three-dimensional streaming structure with two symmetry planes which are parallel and perpendicular to the channel depth plane at the center of the bubble. Each region was shown to contain a toroidal vortex structure, with flow reversal near the ceiling or floor of the channel. This three-dimensional flow structure was modeled by \citet{rallabandi2015three} using a superposition of Stokes solutions to approximate the effects of lateral confinement. \citet{volk2020size} showed that such streaming flows are able to trap particles in three dimensions based on their size, suggesting opportunities for size-dependent particle trapping sorting and focusing applications. While past studies have focused on relatively tall channels, many microfluidic devices utilize much narrower channels, where a systematic theory is missing. 

In this article, we theoretically and experimentally study the effects of vertical confinement on streaming flows, focusing on the limit where vertical length scales are much smaller than lateral scales. We develop a lubrication-like theory describing three-dimensional streaming around cylindrical obstacles with  radius much greater than the depth of cell. Under the Hele--Shaw limit, our theory predicts that the flow reverses direction across the depth of the channel which, to the best of our knowledge, is a feature not observed in less confined microchannels. We show that this feature is confirmed by our experiments of streaming around cylinders sandwiched in a microchannel. In addition, our theory also predicts that the streaming velocity decays as the inverse cube of the distance from the cylinder, faster than that expected from previous two-dimensional approaches. We verify this decay rate quantitatively using particle tracking measurements from experiments of streaming around cylinders with different aspect ratios at different driving frequencies.

\section{Model setup}

We develop a theory for streaming flows in which the characteristic length scale along one dimension (the height of the channel) is much shorter than those along the other two dimensions. Here, we adapt ideas from lubrication theory to the  perturbation framework used to analyze streaming flows generated by small-amplitude oscillations with inertia \citep{riley2001steady}. The resulting theory closely follows the ideas set out by \citet{shih1970unsteady}, although we arrive at different results and conclusions, as we later discuss.

\begin{figure}
   \begin{subfigure}{0.48\textwidth}
       \includegraphics[width=\linewidth]{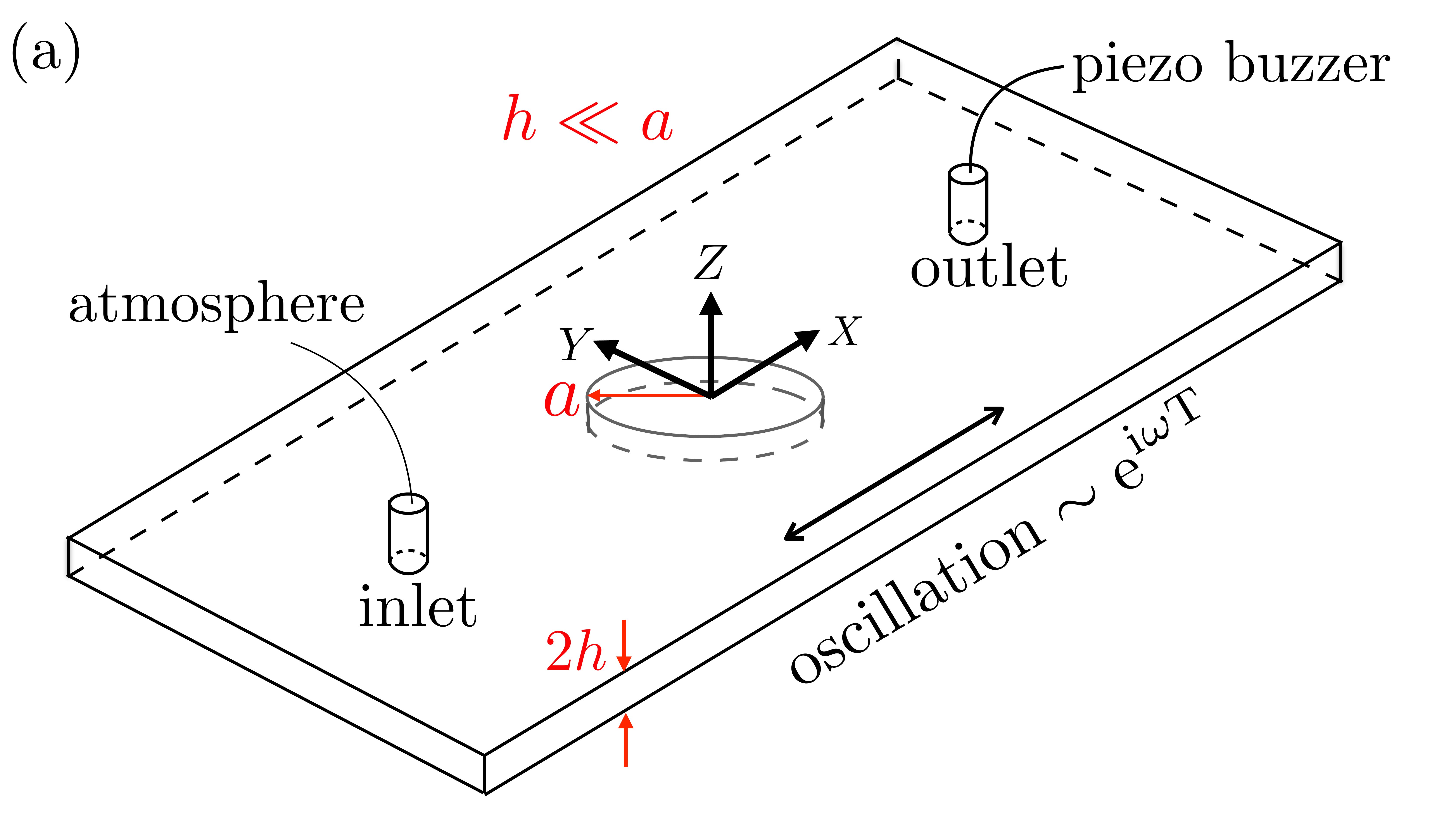}
   \end{subfigure}
\hfill 
   \begin{subfigure}{0.48\textwidth}
       \includegraphics[width=\linewidth]{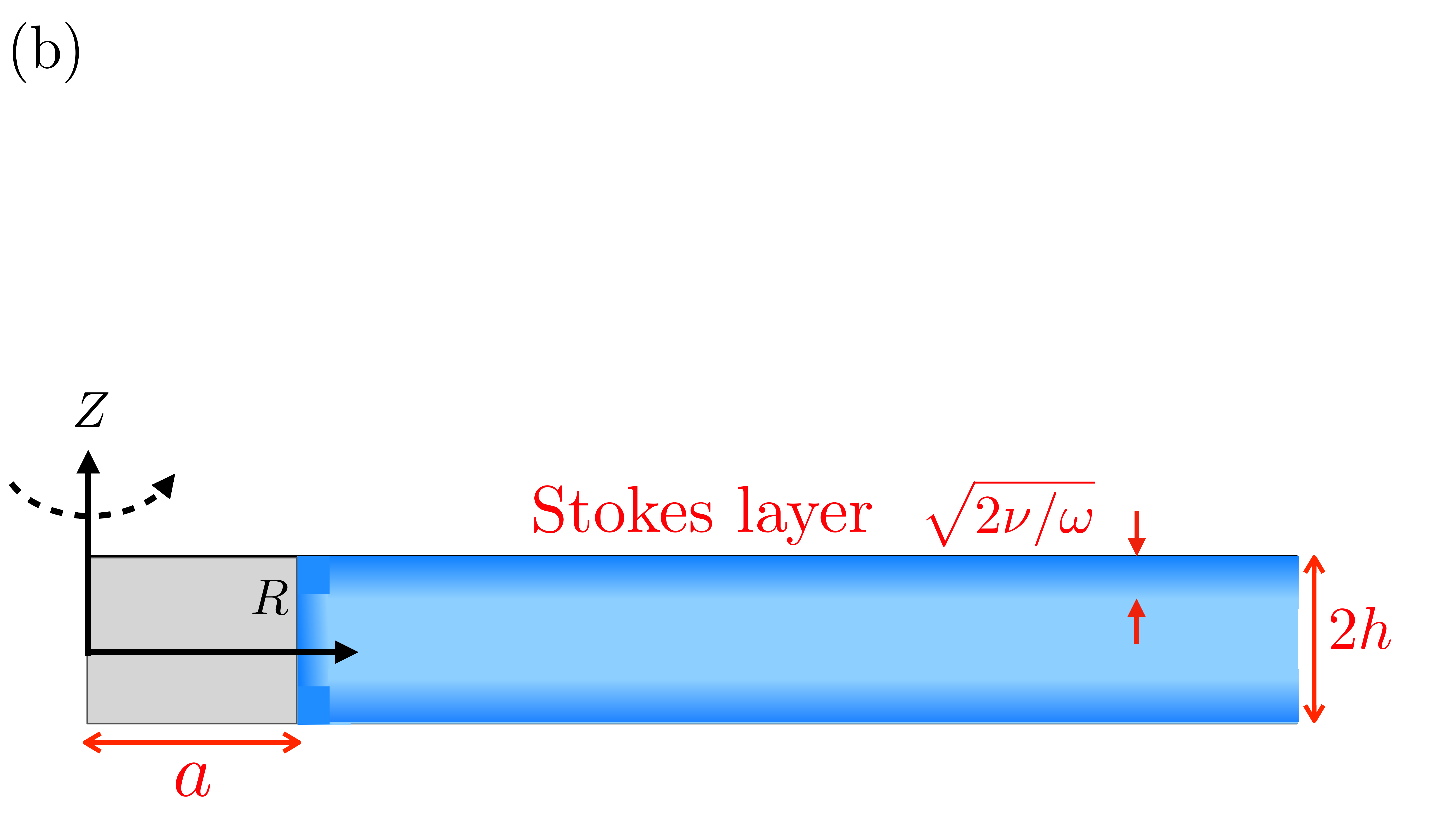}
   \end{subfigure}
   \caption{Model setup: (a) A cylinder sandwiched in the cell with a radius $a$ larger than the height of the cell $2h$. Oscillations of the fluid with angular frequency $\omega$ are driven by a piezo-buzzer. (b) Side view, showing Stokes layers with thickness $\sqrt{2 \nu/\omega}$ near the walls and cylinder surface.}
   \label{fig:model}
\end{figure}

The setup of the theory mirrors the experimental configuration that we  discuss in detail in section \ref{sec:exp}. We consider a cylindrical obstacle that is located in the center of a microfluidic channel with one inlet and one outlet on the top wall (Figure \ref{fig:model}). While the inlet stays exposed to the atmosphere, the outlet is connected to a piezo-buzzer that oscillates the fluid in the channel. The speed of the oscillatory flow is determined by the voltage applied to the buzzer and the angular frequency of oscillation $\omega=2\pi f$. We will assume that the channel's width and length are much greater than both the radius of the cylinder $a$ and the height of the channel $2h$, so that the lateral walls have little influence on the flow.  As a non-trivial restriction, we also assume that the radius of the cylinder is much greater than the half-height of the channel ($a\gg h$).

We describe the flow around the cylinder by introducing a (dimensional) coordinate system $(X,Y,Z)$ centered at the geometric center of the cylinder and oriented as shown in the figure \ref{fig:model}: \textit{X} is along the length of channel, \textit{Y} is across the width of channel and $Z$ is along the depth of channel.  We also define a cylindrical coordinate system $(R,\theta,Z)$ such that $X = R \cos\theta$ and $Y = R \sin\theta$. 
The buzzer applies pressure oscillations to drive an oscillatory flow through the channel. The channel-height-averaged fluid velocity, $ \overline{\bm{V}} =1/(2h) \int_{-h}^h \bm{V}\,\mbox{d}z$ far from the cylinder is directed along the $X$ axis (unit vector $\bm{e}_{X}$) and satisfies 
\begin{equation} \label{pressureBCinf}
    \overline{\bm{V}} (R\rightarrow\infty)\sim \overline{V}_{\infty} \bm{e}_{X} \mathrm{e}^{\mathrm{i}\omega T},
\end{equation}
where $\overline{V}_{\infty}$ is assumed known, and $T$ represents dimensional time. Here and below, it will be understood that only the real part of any complex equation is physically relevant.  Note that the condition $R \to \infty$ refers to distances far from the cylinder that are not too close to the buzzer. 

\subsection{Governing equation and boundary conditions}

The flow satisfies the incompressible Navier-Stokes equations,
\begin{equation}\label{dimNS}
    \rho\left(\frac{\partial\bm{V}}{\partial T}+\bm{V}\cdot \nabla\bm{V}\right)=-\nabla P+\mu\nabla^2\bm{V}, \quad\nabla\cdot\bm{V}=0,
\end{equation}
where \textbf{V}$(\bm{X},T)$ is the three-dimensional (3D) fluid velocity and $P(\bm{X},T)$ is the pressure. The flow satisfies the no-slip condition ($\bm{V=0}$) at the cylinder surface and at the channel walls.

We non-dimensionalize the equations by choosing \textit{a} as the characteristic length scale in the \textit{XY} plane and $h$ as the characteristic scale in the $Z$ direction. We define dimensionless coordinates (lowercase) according to
\begin{align}\label{nondimcoordi}
    X=ax,\quad Y=ay,\quad Z=hz.
\end{align}
For later convenience, we write the velocity as $\bm{V} = \bm{V}_{\parallel} + V_Z \bm{e}_Z$, where $\bm{V}_{\parallel}$ represents velocities in the horizontal ($XY$) plane and $V_Z$ is the vertical velocity component. The velocity scale $\overline{V}_\infty$ is characteristic of the horizontal velocity $\bm{V}_{\parallel}$. Continuity then suggests that a vertical velocity scale $\overline{V}_\infty h/a$. Substituting these velocity scales into \eqref{dimNS} and comparing viscous and pressure terms suggests a pressure scale $\mu \overline{V}_{\infty} a/h^2$. We therefore define the dimensionless velocity components $\bm{v}_{\parallel}$ and $v_z$ and pressure $p$ according to
\begin{equation}\label{nondimpuv}
    \quad\bm{V}_{\parallel}= \overline{V}_\infty \bm{v}_{\parallel},\quad V_Z=\frac{\overline{V}_\infty h}{a} v_z, \quad  P= \frac{\mu \overline{V}_\infty a}{h^2} p.
\end{equation}
We also define a rescaled time $t =  \omega T$ and introduce the dimensionless parameters, 
\begin{equation}\label{nondimpara}
    \delta = \sqrt{\frac{2\nu}{h^2 \omega}},\quad\epsilon=\frac{\overline{V}_{\infty}}{a \omega},\quad \mbox{and} \quad \eta=\frac{h}{a}.
\end{equation}
Here, $\epsilon$ is the dimensionless amplitude of the oscillatory flow, $\delta$ is the ratio of Stokes layer thickness to the channel height and characterizes the frequency of the flow, while $\eta$ is the aspect ratio of the cylinder and characterizes the degree of axial confinement.

The 3D fluid velocity is therefore $\bm{v} = \bm{V}/\overline{V}^{\infty} = \bm{v}_{\parallel} + \eta v_z \bm{e}_z$. Substituting for all the quantities in terms of dimensionless variables into \eqref{dimNS} yields
\begin{subequations}\label{nondimNS}
\begin{align}
    &\frac{2}{\delta^2} \left(\frac{\partial \bm{v}_{\parallel}}{\partial t}+\epsilon\,\bm{v} \cdot \nabla\bm{v}_{\parallel}\right)= -\nabla_{\parallel} p+\left(\eta^2\,\nabla^2_{\parallel} +\frac{\partial ^2}{\partial z^2}\right) \bm{v}_{\parallel},\\
    &\frac{2}{\delta^2} \left(\frac{\partial v_z}{\partial t}+\epsilon\,\bm{v} \cdot \nabla v_z\right)= -\frac{1}{\eta^2} \frac{\partial p}{\partial z}+\left(\eta^2\,\nabla^2_{\parallel}+\frac{\partial ^2}{\partial z^2}\right)v_z,\\
    &\nabla_{\parallel}\cdot\bm{v}_{\parallel}+\frac{\partial v_z}{\partial z}=0,
\end{align}
\end{subequations}
where $\nabla_{\parallel}=\bm{e}_{x}\frac{\partial}{\partial x}+\bm{e}_{y}\frac{\partial}{\partial y}$ is the gradient in the horizontal plane. Equation \eqref{nondimNS} is an exact dimensionless version of \eqref{dimNS} without approximations. The height-averaged velocity at infinity becomes $\overline{\bm{v}}(r \rightarrow\infty)=\bm{e}_{x}\mathrm{e^{it}}$, and the flow satisfies no-slip conditions on the cylinder surface and channel walls.

\section{Streaming theory for narrow channels}\label{section3}

We focus on the limit where the channel is narrow ($h \ll a$), so that $\eta \ll 1$. In this limit, the domain of the flow can be split into an outer region where distances from the cylinder surface are much greater than the channel height ($r - 1 \gg \eta$) and an inner region of thickness comparable to the channel height surrounding the cylinder surface ($r - 1 = O(\eta)$). As we show below, the flow in the outer region can only approximately satisfy the boundary conditions on the cylinder. The role of the inner region is to compensate for the outer flow and meet the no-slip condition on the cylinder surface. Here, we consider the limit $\eta \ll 1$ (so the inner region is asymptotically thin) and focus mainly on the flow in the outer region, although we discuss some details of the inner region in Appendix \ref{appA}.

In the outer region, viscous stresses are dominated by shear across the height of the channel, and equation \eqref{nondimNS} approximates to 
\begin{subequations}\label{simnondimNS}
\begin{align}
    &\frac{2}{\delta^2}\left(\frac{\partial \bm{v}_{\parallel}}{\partial t}+\epsilon\,\bm{v} \cdot \nabla_{\parallel}\bm{v}_{\parallel}\right) = -\nabla_{\parallel} p+\frac{\partial^2\bm{v}_{\parallel}}{\partial z^2},\label{simnondimNSa}\\
    &\frac{\partial p}{\partial z}=0,\\
    &\nabla_{\parallel}\cdot\bm{v}_{\parallel}+\frac{\partial v_z}{\partial z}=0.
\end{align}
\end{subequations}
The first equation is the approximate momentum equation in the $xy$ plane, the second is the approximate momentum equation in $z$ and the third is continuity. For small $\eta$, we find that the outer flow satisfies the effective boundary condition of zero channel-averaged normal velocity normal to the cylinder surface, up to corrections of $O(\eta)$; see Appendix \ref{appA}. Denoting the unit vector normal to the cylinder surface by $\bm{n}$ and using overbars to indicate the average across the height of the channel ($\overline{g} \equiv 1/2 \int_{-1}^1g(z)\,\mbox{d}z$), (\ref{simnondimNS}) are therefore subject to the boundary conditions
\begin{subequations}\label{nondimNSbc}
\begin{align}
    &\overline{\bm{v}} = \bm{e}_x \mathrm{e^{it}} , \quad \mbox{as} \quad r\rightarrow\infty,\\
    &\bm{v} =\bm{0},\quad \mbox{at} \quad z=\pm 1\quad \mbox{(top and bottom walls)},\\
    &\overline{\bm{v}}\cdot \bm{n}=0, \quad \mbox{at} \quad r=1 \mbox{(cylinder surface)}.\label{zerochannelaverage}
\end{align}
\end{subequations}
Here, $\overline{\bm{v}}$ is the mean velocity across the channel height. 

\subsection{Small amplitude approximation and result for outer flow}

Equations \eqref{simnondimNS} and \eqref{nondimNSbc} are the simplified forms for a thin channel. We solve these equations approximately in the practically relevant limit of $\epsilon\ll 1$ and with arbitrary $\delta$. For small $\epsilon$, we develop a solution by applying a perturbation expansion \citep{riley2001steady}
\begin{align}\label{seriesexpan}
    (p, \bm{v})=(p_1, \bm{v}_1)+\epsilon (p_2, \bm{v}_2)+\cdots,
\end{align}
The subscript $1$ identifies the primary flow (oscillatory flow) whereas the subsript $2$ corresponds to the secondary steady flow (streaming). The above series expansion, when substituted into \eqref{simnondimNS} and separating orders of $\epsilon$, yields the governing equations for the primary and secondary flow.

\subsubsection{Primary flow}

The governing equations with boundary conditions for the primary flow at $O(\epsilon^0)$ are 
\begin{align}\label{prifloweq}
    \frac{2}{\delta^2}\frac{\partial \bm{v}_{1 \parallel}}{\partial t}&= -\nabla p_1+\frac{\partial^2 \bm{v}_{1 \parallel}}{\partial z^2}, \quad \frac{\partial p_1}{\partial z}=0, \quad \nabla\cdot\bm{v}_{1 \parallel}+\frac{\partial v_{z1}}{\partial z}=0,\quad\mbox{with}\nonumber\\
    \quad \overline{\bm{v}}_1|_{r\rightarrow\infty} &=\overline{\bm{e}}_{x}\mathrm{e^{it}},\quad \bm{v}_{1 \parallel}|_{z=\pm 1}=\bm{0}, \quad v_{z1}|_{z=\pm 1}=0,\quad \overline{\bm{v}}_{1 \parallel }\cdot \bm{n}|_{r=1}=0.
\end{align}
The system \eqref{prifloweq} is linear and the only non-homogeneous condition is the velocity condition at infinity. We therefore seek separable solutions proportional to $\mathrm{e^{it}}$. Substituting this ansatz into \eqref{prifloweq} and observing that the pressure is independent of $z$,  the solution to the primary flow is 
\begin{subequations}\label{priflowsolu}
\begin{align}
    p_1&= G(\alpha)\left(r+\frac{1}{r}\right)\cos{\theta}\,\mathrm{e^{it}}, \quad\mbox{where}\quad G(\alpha) =-\frac{\alpha^3}{\alpha-\tanh{(\alpha})}, \quad 
    \alpha=\frac{(1+\mathrm{i})}{\delta}, \label{pripressure}\\ 
    \bm{v}_1&=-\frac{\nabla p_1}{\alpha^2}\left(1-\frac{\cosh{\alpha z}}{\cosh{\alpha}}\right)\mathrm{e^{it}}.\label{eqnv1}
\end{align}
\end{subequations}
Observe that the velocity of primary flow depends on Stokes layer thickness and the primary pressure involves $r$ and $\theta$. There is no vertical velocity ($v_{z1}=0$). The solution for primary flow is in agreement with that of \citet{shih1970unsteady}. The channel-averaged oscillatory velocity is therefore 
\begin{align}\label{v1k}
    \overline{\bm{v}}_1 = \frac{\nabla p_1}{G(\alpha)} = \nabla \left[\left(r+\frac{1}{r}\right)\cos{\theta}\,\right] \mathrm{e^{it}}, 
\end{align}
which is $O(1)$ by definition. We will use \eqref{v1k} in the calculation of the streaming flow below. 
 
\subsubsection{Secondary flow}\label{sec:secondary solution}

The convective acceleration term in \eqref{simnondimNSa}, which is neglected when solving for the primary flow, appears as a body force that drives the secondary flow. It will be understood that all secondary quantities are time-averaged. The governing equations at $O(\epsilon^1)$ are 
\begin{subequations}\label{secfloweq}
\begin{align}
    \frac{2}{\delta^2}\left<\bm{v}_{1 \parallel}\cdot\nabla\bm{v}_{1 \parallel}\right>&= -\nabla p_2+\frac{\partial^2 \bm{v}_2}{\partial z^2}, \quad \frac{\partial p_2}{\partial z}=0, \quad \nabla\cdot\bm{v}_{2 \parallel}+\frac{\partial v_{z2}}{\partial z}=0, \quad\mbox{with} \\
    \quad p_2|_{r\rightarrow\infty} &= 0,\quad \bm{v}_{2 \parallel}|_{z=\pm 1}=\bm{0}, \quad v_{z2}|_{z=\pm 1}=0,\quad\overline{\bm{v}}_{2 \parallel }\cdot \bm{n}|_{r=1}=0.\label{secflowbc}
\end{align}
\end{subequations}
Substituting \eqref{eqnv1} into the advective term $\left<\bm{v}_{1 \parallel}\cdot\nabla\bm{v}_{1 \parallel}\right>$ and integrating the momentum equation, we obtain:
\begin{align}\label{generalv2}
    \bm{v}_{2 \parallel}=&\,\frac{1}{2}(z^2-1)\,\nabla p_2+ g_1(z,\alpha)\nabla|\nabla p_1|^2,\quad\mbox{where}\\
    g_1(z,\alpha)=&\frac{\mathrm{i}}{8\alpha^4}\left((z^2-1)\alpha^2+\frac{2\cos{\alpha z}}{\cos \alpha}-\frac{2\cosh\alpha z}{\cosh\alpha}+\frac{\sin\alpha z\,\sinh\alpha z}{\cos\alpha\,\cosh\alpha}-\tan\alpha\, \tanh\alpha\right)\nonumber.
\end{align}
Integrating the continuity equation in \eqref{secfloweq} and applying the boundary conditions at the top and bottom walls, we find that $\nabla\cdot\overline{\bm{v}}_{2 \parallel }=0$ (the horizontal flux is divergence-free).  Substituting \eqref{generalv2} into the above condition, we find that $p_2$ is governed by the Poisson equation $\nabla_{\parallel}^2 {p_2}= 3\,\overline{g}_1\nabla_{ \parallel}^2|\nabla p_1|^2,$ which relates it to the inertia of the primary flow. 

The solution to the secondary pressure can be decomposed into a homogeneous solution $p_2^{hom}$ (which satisfies Laplace's equation) and a particular solution, which is simply $3\,\overline{g}_1|\nabla p_1|^2$. Requiring that (i) the secondary pressure decays far away from the cylinder and (ii) the channel-height-averaged secondary normal velocity at the cylinder surface vanishes (see \eqref{secflowbc}), we find that the homogeneous solution is equal to zero. Thus, we finally obtain the secondary pressure as 
\begin{equation}\label{equp2}
    p_2=-\frac{\mathrm{i}}{16\alpha^5}\left(4 \alpha^3 + 15 (\tanh \alpha - \tan \alpha)  +6\alpha\tan\alpha\,\tanh{\alpha}\right)|\nabla p_1|^2.
\end{equation}
Substituting the expression above into (\ref{generalv2}) yields $\bm{v}_{2 \parallel}$ and using the continuity equation yields the vertical component of secondary flow $v_{z2}$. Observing that the secondary flow pressure and velocity both contain the term $|\nabla p_1|^2=\left| G(\alpha) \right|^2 |\overline{\bm{v}}_1|^2$ (see \ref{v1k}), we obtain for the secondary (time-averaged) velocity
\begin{subequations}\label{secflowsolu}
\begin{align}
    \bm{v}_2 &= \bm{v}_{2 \parallel} + \eta v_{2 z} \bm{e}_z = F_{\parallel}(z,\alpha)\;\nabla\left(|\overline{\bm{v}}_1|^2 \right) + \eta F_z(z, \alpha)\nabla^2\left(|\overline{\bm{v}}_1|^2 \right) \bm{e}_z, \quad \mbox{where} \\
   F_{\parallel}(z,\alpha)&=\frac{\mathrm{i} \left| G(\alpha) \right|^2}{32\alpha^5}\biggl(\frac{8\alpha\cos{\alpha z}}{\cos{\alpha}}-\frac{8\alpha\cosh{\alpha z}}{\cosh{\alpha}}+\frac{4\alpha\sin{\alpha z}\,\sinh{\alpha z}}{\cos{\alpha}\,\cosh{\alpha}}\nonumber\\
    &\qquad+15(z^2-1) \left(\tan \alpha -\tanh \alpha\right)- 2(3 z^2-1) \alpha\tan\alpha\, \tanh{\alpha}\biggr),\\
  F_z(z,\alpha)&=-\frac{\mathrm{i} \left| G(\alpha) \right|^2}{32\alpha^5}\biggl(\frac{8\sin{\alpha z}}{\cos{\alpha}}- \frac{8\sinh{\alpha z}}{\cosh{\alpha}} +\frac{2\left(\sin{\alpha z}\,\cosh{\alpha z}- \cos\alpha z\, \sinh{\alpha z}\right)}{\cos{\alpha}\,\cosh{\alpha}}\nonumber\\
    &\qquad+5z(z^2-3)(\tan \alpha- \tanh \alpha) - 2 z (z^2-1) \alpha \tan \alpha \tanh \alpha\biggr). 
\end{align}
\end{subequations}
Observe that the functions $F_{\parallel}(z,\alpha)$ and $F_z(z,\alpha)$ describe horizontal and vertical velocity profiles across the height of the channel, while horizontal variations are captured by $|\overline{\bm{v}}_1|^2$. 

While $\bm{v}_2$ describes the Eulerian secondary flow,  it is the time-averaged  motion of material fluid elements (the Lagrangian secondary flow) that is practically relevant and is normally referred to as streaming. The time-averaged Lagrangian velocity is $\bm{v}_L=\bm{v}_2+\bm{v}_d$, where $\bm{v}_d$ is the Stokes drift defined by $\bm{v}_d=\left<(\int\bm{v}_1\,\mbox{d}t\cdot\nabla)\bm{v}_1\right>$ \citep{riley2001steady}. Under the framework developed here, the Stokes drift is related to the primary pressure by 
\begin{align}\label{stokedrift}
    \bm{v}_d =  \frac{1}{2}\left|\frac{1}{\alpha^2}\left(1-\frac{\cosh{\alpha z}}{\cosh{\alpha}}\right)\right|^2 (\mathrm{i} \nabla p_1^* \cdot \nabla \nabla p_1),
\end{align}
where the asterisk denotes the complex conjugate. For the $p_1$ given in \eqref{pripressure}, the Stokes drift vanishes identically, so the Lagrangian streaming is given simply by $\bm{v}_2$.

Because the homogeneous part of $p_2$ and the Stokes drift both vanish to leading order in $\eta$, the streaming velocity has zero average across the channel \emph{everywhere} in the flow, not just at the cylinder surface (i.e. $\overline{F}_{\parallel} = 0$). We argue in appendix  \ref{appA} that the leading contribution to the channel-averaged streaming occurs at $O(\eta)$ through details of the inner region. 

\subsection{Flow structure}

\begin{figure}
        \includegraphics[width=1\linewidth]{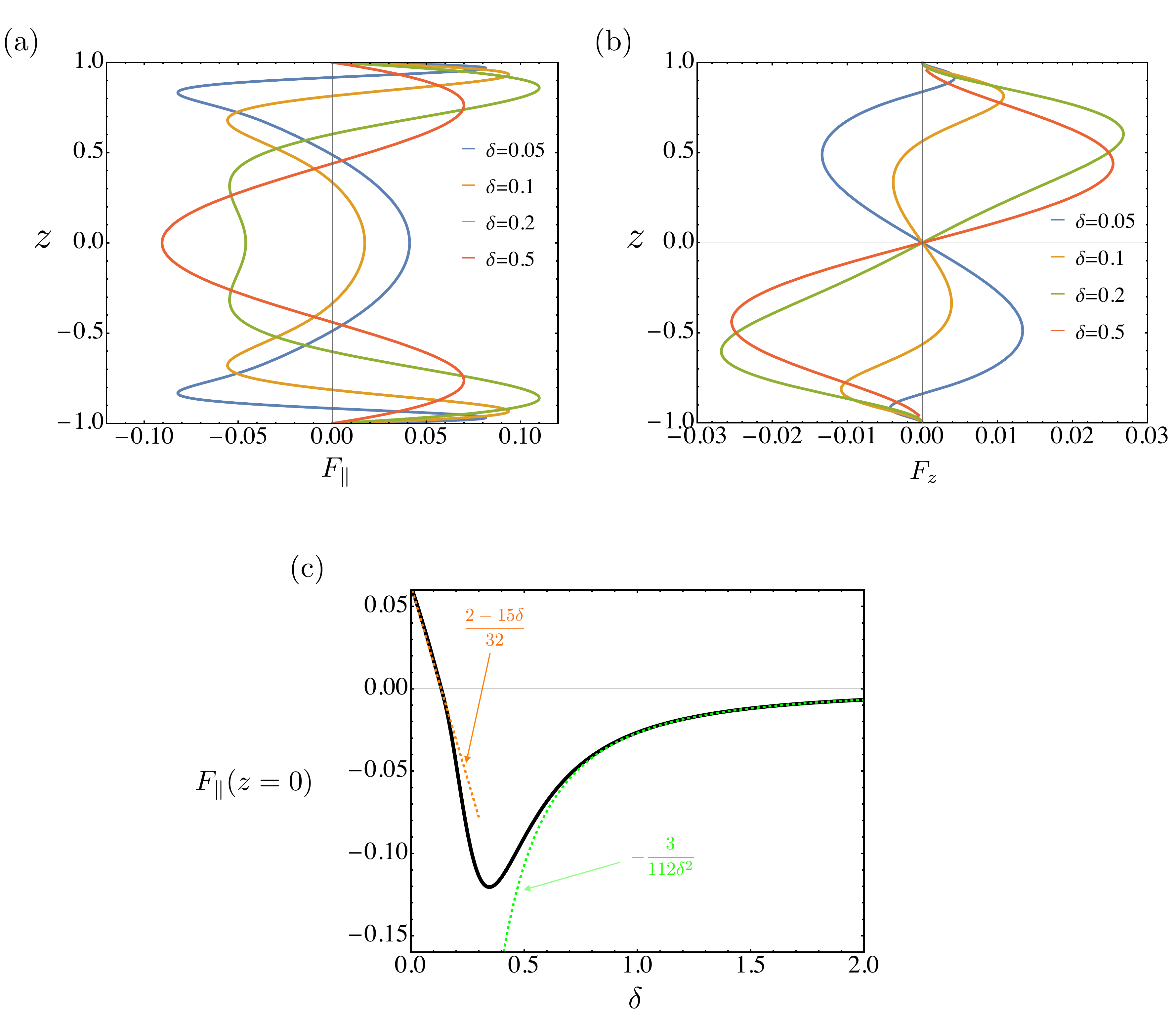}
        \caption{Analytical description of streaming velocity: (a) horizontal and (b) vertical velocity profiles for different values of Stokes layer thickness $\delta$. The flow reverses direction multiple times across the height of the channel depending on the value of $\delta$. (c) horizontal velocity at the central plane $z=0$ reverses direction across $\delta \simeq 0.13$. Dashed curves are asymptotic expansions for small and large $\delta$.} 
\label{fig:v1}
\end{figure}

The variation of the streaming velocity across the height of the channel, characterized by the functions $F_{\parallel}$ and $F_z$, is shown in Figure \ref{fig:v1} for different values of $\delta$. Horizontal velocity profiles for small $\delta \lesssim 0.1$ have similar shapes. Near the top and bottom walls, the horizontal flow increases and decreases rapidly on the scale of the Stokes layer thickness $\delta$ before changing direction (figure \ref{fig:v1}(a)). At the outer edge of the Stokes layers the flow achieves a maximal velocity (the maximal negative value of $F_{\parallel}$), which we identify with the time-averaged slip over walls at the edge of thin Stokes layers \citep{longuet1953mass,nyborg1958acoustic}. The central core outside the Stokes layers is characterized by a more gradual variation of the flow, which reverses direction another time as one moves towards the central  plane ($z=0$) of the channel. For $\delta \gtrsim 0.13$, the second flow reversal near $z = 0$ disappears (see profile for $\delta = 0.2$ in figure \ref{fig:v1}(a)). For $\delta$ larger than about $0.25$, the shape of flow remains roughly the same and is characterized by a single flow reversal with a maximum at the central plane where $z=0$. For $\delta \ll 1$, the central core (outside Stokes layers) is described by the parabolic profile with slip $F_{\parallel} \sim 1/16(1-3z^2)$, whereas for $\delta \gg 1$, the flow profile is described by  $F_{\parallel} \sim (3 \delta^{-2}/560) (z^2 - 1)(7 z^4 - 28 z^2 +5)$. 

The vertical velocity, shown in figure \ref{fig:v1}(b), exhibits odd symmetry about $z = 0$ but undergoes similar changes in sign as $\delta$ increases. 
When $\delta$ is small, $F_z$ is mostly negative in the top half of the channel with a reversal in direction within the Stokes layer. By contrast, for $\delta$ larger than about $0.13$, the vertical velocity reverses and $F_z$ is positive throughout the top half of the channel. 

Figure \ref{fig:v1}(c) shows the behavior of the horizontal velocity at the center plane  $z=0$. As $\delta$ increases, the horizontal velocity decreases and then increases rapidly to reach a maximum (at about $\delta=0.35$), with a direction change occurring at about $\delta \simeq 0.13$. The orange dashed curve represents asymptotic expansions of horizontal center-plane velocity function $F_{\parallel}(z = 0) \sim (2-15\delta)/32$ for small $\delta$, whereas the green dashed curve shows the behavior of $F_{\parallel}(z = 0) \sim -3/112\delta^2$ for large $\delta$. 

\begin{figure}
\includegraphics[width=1\textwidth]{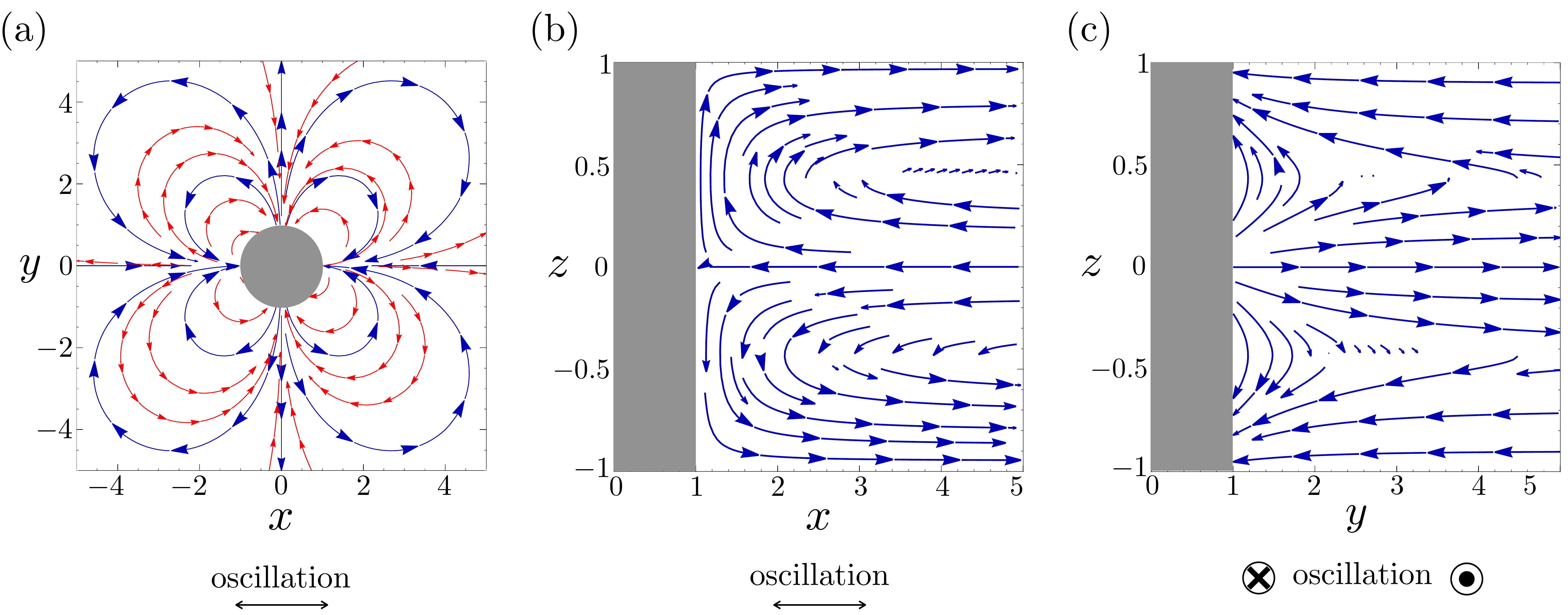} 
   \caption{Theoretical streamline plots at $\delta=1$, with oscillations driven along $x$. (a) Streamlines in the horizontal plane show a vortex structure that reverses direction across $z$ (blue and red streamlines are at $z = 0$ and $z = 0.5$, respectively). The reversal of flow is made clearer by streamlines in the vertical planes (b) and (c).}
   \label{fig:StreamlinePlot}
\end{figure}

In addition to showing velocity distributions, we visualize the  three-dimensional flow using streamline portraits in two-dimensional sections. Streamline plots in different planes `length-width\,($\emph{xy}$)', `length-depth\,($\emph{xz})$' and `width-depth\,($\emph{yz}$)' planes are shown in Figure \ref{fig:StreamlinePlot} with the oscillation direction and coordinates marked out.

Figure \ref{fig:StreamlinePlot}(a) shows streamlines around a cylinder (shaded gray) on the horizontal plane but at different depths at $\delta = 1$: Blue streamlines represent the flow at the axial mid-plane ($z=0$), showing a  qualitatively similar structure to two-dimensional streaming. However, streamlines at a depth $z=0.5$, shown in the same figure in red, indicate flow in the opposite direction. This reversal of flow direction (associated with changes in the sign of $F_{\parallel}$) is a feature that is clearly absent from two-dimensional theory that ignores the depth of cells. It is interesting to note that previous experimental and theoretical descriptions of three-dimensional streaming flows under less extreme vertical confinement (of rigid cylinders and semi-cylindrical bubbles) also do not report such a flow reversal \cite{marin2015three,rallabandi2015three}. In the following section, we present direct experimental evidence of axial flow reversal for highly confined streaming setups. Figures \ref{fig:StreamlinePlot}(b) and \ref{fig:StreamlinePlot}(c) indicate flows in vertical planes and also show that the direction of secondary flow reverses across the depth of the channel. We recall that the vertical distributions of velocity are qualitatively similar for $\delta \gtrsim 0.25$ (cf. Figure \ref{fig:v1} and surrounding discussion). 

We observe in figure \ref{fig:StreamlinePlot}(c) in the `width-depth' plane that streamlines terminate at $y=1$ at the cylinder boundary. Unlike figure \ref{fig:StreamlinePlot}(b), we discover that the flow appears to locally penetrate the cylinder surface, even though the depth-averaged velocity normal to the cylinder surface remains identically zero by construction. This is understood by recalling that the present theory is restricted to the ``outer region'' where $r - 1 \gg \eta$. To accommodate the local velocity normal to the cylinder at $r=1$, it is necessary to acknowledge the presence of an inner region of width $\eta$ around the cylinder that ensures that the streaming velocity vanishes everywhere on the cylinder surface. Though we do not resolve the details of this inner region here, we provide some estimates of its effects in Appendix \ref{appA}.

\section{Experiments}\label{sec:exp}

We show that experiments for streaming around a cylinder sandwiched in a Hele-Shaw channel (narrow height) support our theoretical prediction that the direction of flow reverses across the depth of cell.

We fabricate devices (see figure \ref{fig:model}) out of polydimethyisiloxane (PDMS, Dow Sylgard 184) to obtain transparent microchannels with height $2h$ at 160 $\mu$m or 480 $\mu$m, containing cylinders with radius $a$ of 600 $\mu$m or 750 $\mu$m inside. The PDMS mixture is poured onto molds (made from layers of backing tape) and peeled after curing on a hot plate for about half an hour. We mount a microchannel securely onto a movable microscope stage that can be controlled manually. Deionized water ($\nu = 10^{-6}$ m$^2$/s, density 1.00 g/cm$^3$) with polystyrene microparticles (Magsphere; diameter $2 a_p = 5.0$ $\mu $m, density 1.05  g/cm$^3$) is injected through tubing into the microchannel using a syringe. The injection is stopped once the microchannel is full and syringe is disconnected, exposing the inlet to atmosphere (see figure \ref{fig:model}). The piezo-buzzer connected to the other end (outlet) is then turned on, and is driven by applying sinusoidal signals of various frequencies from a function generator (5V, Rigol DG4062) amplified through a power amplifier (Krohn-hite 7500). The buzzer produces pressure oscillations in the channel to drive oscillatory flow which in turn results in streaming. A high-speed camera (Photron, Fastcam Nova S6) captures images or videos through an inverted microscope (Leica DMi8) from below with a choice of 4x and 10x microscope objective lenses. The frame rate is chosen to be a integer divisor of the driving frequency  so as to isolate the steady component of the particle motion. 

\subsection{Visualization}

\begin{figure}
\centering
\includegraphics[width=1\textwidth]{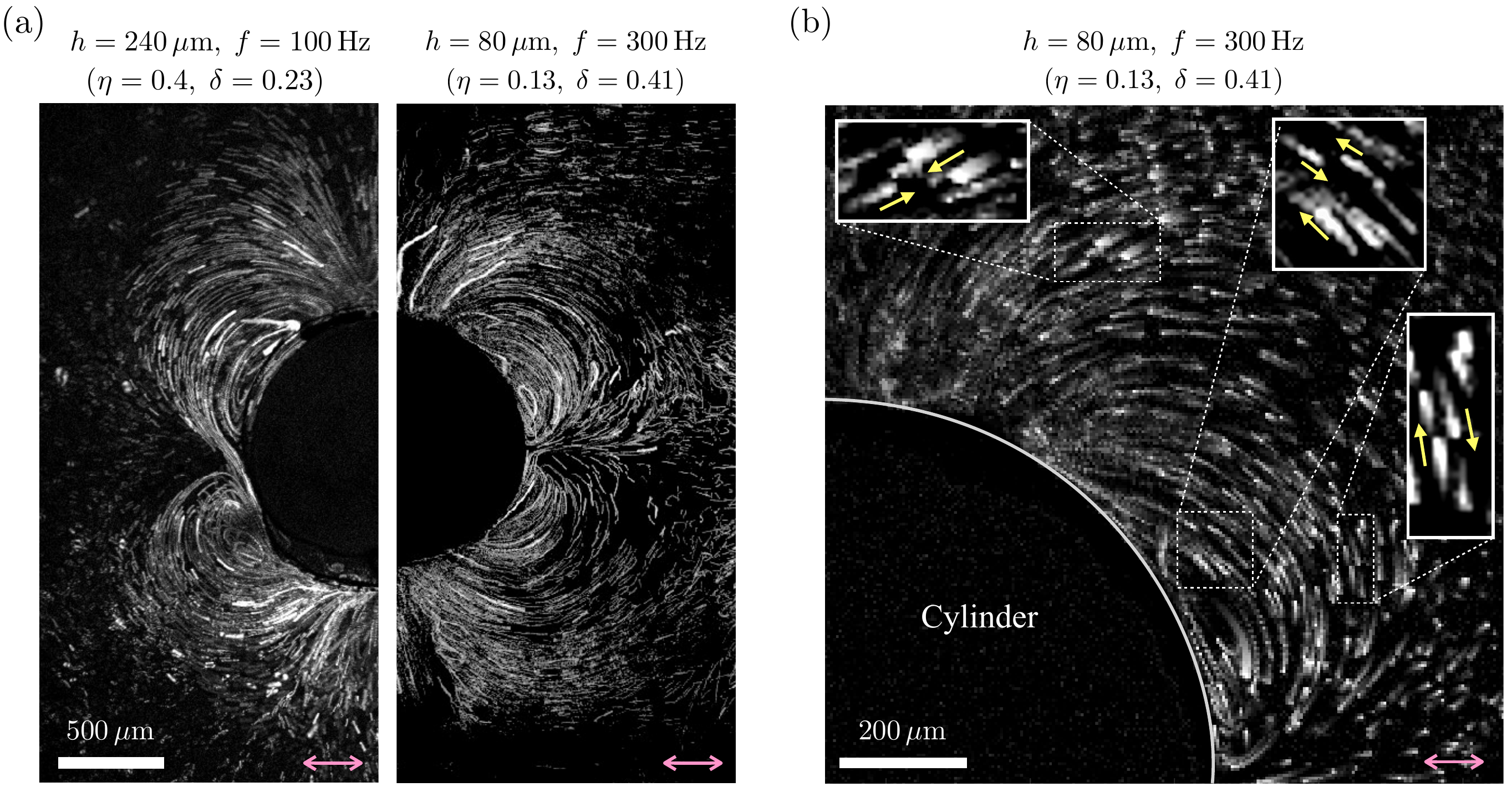}
   \caption{Pathlines of tracer particles around a cylinder of radius $a = 600\,\mu$m for different experimental conditions; double arrows indicate the direction oscillation. (a) Steady pathlines for two experiments with different channel height $h$ and frequency $f$. The qualitative flow pattern is robust across our experiments. (b) Particle traces for $\eta = 0.13$ (corresponding to flow in the top half of the right panel of (a)), but now including temporal information: For each trace, the brighter ``head" shows the particle's future position, whereas the darker ``tail'' represents the particle's past position. The insets details of different parts of the flow where pairs of microparticles are travelling in opposite directions,  but at different depths.}
   \label{fig:experiment}
\end{figure}

Matlab and ImageJ are used to process the recorded video to obtain calibrated images of the microparticles as they are transported by the flow. The microparticles are expected to faithfully behave as passive tracers of the flow due to their small size ($a_p/a \lesssim 0.005$) and inertia (Stokes number $a_p^2 \omega/\nu \lesssim 0.01$). We extract the pathline structure from the experimental images by using an in-house implementation of the Flowtrace algorithm \citep{gilpin2017flowtrace}.  Figure \ref{fig:experiment}, with the oscillation direction represented by pink arrows, indicates  steady pathlines of tracers around a cylinder of radius $a = 600\, \mu$m) but with different $h$ and driving frequency $f$. The flow structure is robust across our experiments: Four steady vortices are created adjacent to the cylinder. Figure \ref{fig:experiment}(a) shows these vortices for two representative experiments (one pair of vortices per experiment). The flow structure projected in the imaging plane $(xy)$ resembles two-dimensional streaming around a cylinder in an unbounded or concentrically bounded fluid, and is qualitatively similar to the theoretical streamline portrait of figure \ref{fig:StreamlinePlot}(a). 

A closer look at the trajectories for small $\eta$, however, reveals a  striking difference from two-dimensional streaming. Figure \ref{fig:experiment}(b) shows trajectories of particles in an experiment with $\eta \approx 0.13$ (cf. right panel in figure \ref{fig:experiment}(a)), digitally constructed such that the brighter part of each trace represents the future position of one particle (head of the trace) and the darker part represents the past position of the same particle (tail of the trace); see \cite{gilpin2017flowtrace}. This leads to the appearance of thicker heads and thinner tails of the particle tracers (note that this is not an optical effect but a digital one that accounts for temporal information). We observe pairs of apparently nearby microparticles that are clearly traveling in opposite directions. 
Zooming into one of the marked regions (insets in figure \ref{fig:experiment}(b)) makes the counter-motion of nearby particles clearer; yellow arrows indicate the direction of motion to guide the eye. 
We observe this feature for various experimental conditions, and across the entire field of view (see supplementary movie). 

Tracer particles moving in opposite directions at the same literal locations (identical $x$, $y$ and $z$) would violate continuity. Thus, the observation of counter-moving particle trajectories at the same projected location in the $xy$ plane implies that the trajectories in question must be at different depths $z$. We note that this feature is not localized and can be observed everywhere within the field of view in Figure \ref{fig:experiment}(b), though it is difficult to visualize opposing trajectories everywhere for the same set of images. A much clearer visualization is offered by the high speed movie of the trajectories (Supplementary Information). The movie also makes it clear that particles moving in one sense around the vortex are brighter than particles moving in the opposite sense, which suggests that these sets of particles are at different distances from the focal plane, and thus at different depths $z$. This experimental conclusion is in agreement with the theoretical prediction, see figures \ref{fig:v1} and \ref{fig:StreamlinePlot}. Though we are unable to directly measure depth information, the observation of particle traces along with the continuity argument above offers convincing evidence to support the theoretical prediction that the streaming velocity changes direction across the depth of the channel.

\subsection{Velocity decay}

Our streaming theory predicts a three-dimensional velocity field $\bm{v}_2=\bm{v}_{2 \parallel}+\eta\,v_{z2}\,\Bar{\mathrm{e}}_z$. Substituting $|\overline{\bm{v}}_1|^2=\left(1+1/r^4-2 \cos{(2 \theta)}/r^2\right)$ from \eqref{priflowsolu} into \eqref{secflowsolu} we obtain
\begin{align}
    \bm{v}_2=\left(v_{r2},\, v_{\theta 2},\, \eta\, v_{z 2}\right) = \left(\frac{4}{r^5}(r^2\cos{2\theta}-1) F_{\parallel},\;\frac{4}{r^3}\sin{2\theta}\,F_{\parallel},\;\frac{16 \eta}{r^6}F_z\right)\!.
\end{align}
As is typical for streaming flows, the flow speed decays with the distance from the obstacle. Two-dimensional theory predicts that the velocity decays with the square of the distance to the center of the cylinder $|\bm{v}_2| \propto r^{-2}$. However, observe from the above solution that horizontal velocities are predicted to have a dominant far-field decay $|\bm{v}_2| \propto r^{-3}$. 

\begin{figure}
\centering
\includegraphics[width=1\textwidth]{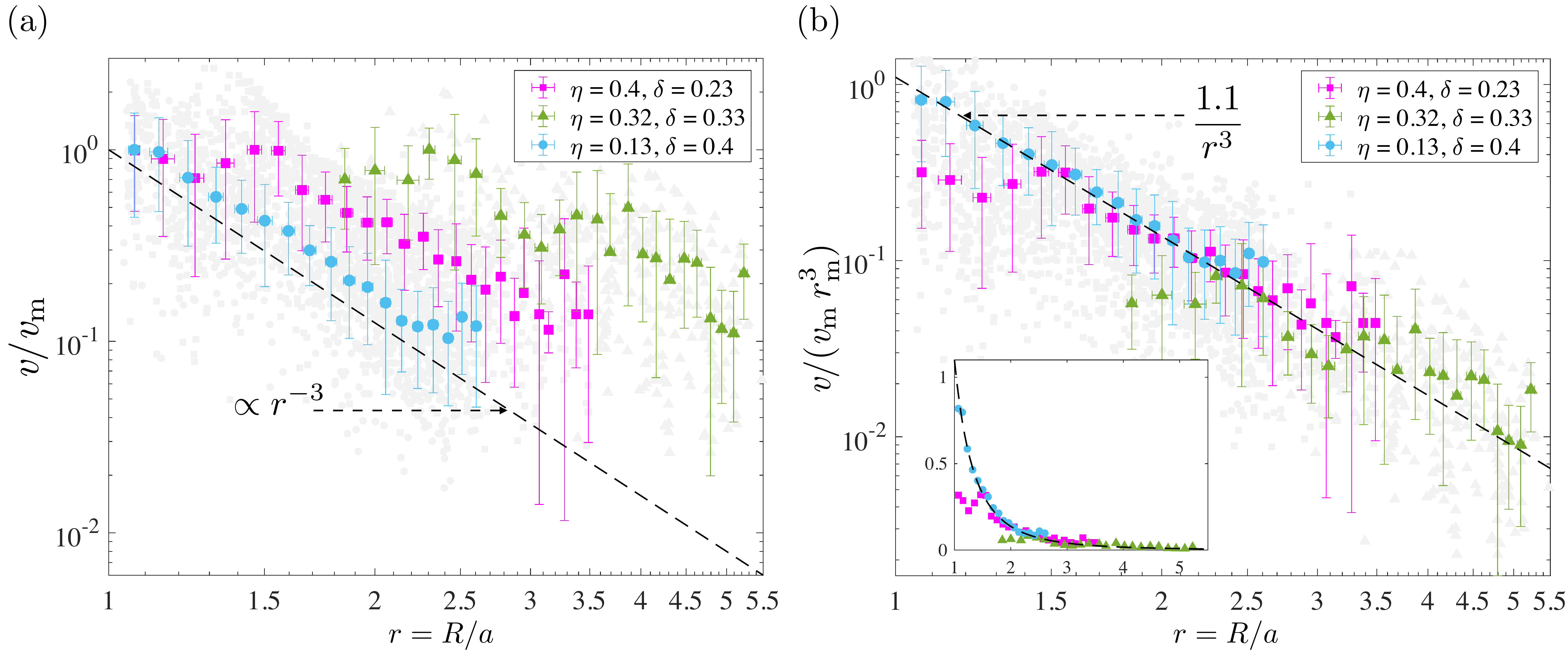} 
   \caption {Decay of normalized horizontal speed $v/v_{\rm m}$ with dimensionless distance from the center of the cylinder $r = R/a$ showing experiments (symbols) and the theoretical prediction (dashed line). The maximal value for each experiment is $v_{m}$. Light gray symbols are velocity data from individual particles at $\theta = \pi/4 \pm 0.1$ whereas colored symbols are data after a local averaging. (a) Data for $v/v_{\rm m}$ from different experiments decay approximately as $r^{-3}$ but do not overlap due to different aspect ratio $\eta$ and Stokes layer thickness $\delta$. 
   (b) Rescaling velocities as suggested by the present theory causes data from different experiments to collapse onto a universal curve $v/(v_{\rm m} r_{\rm m}^3) \approx 1.1 r^{-3}$. The inset shows the same data on a linear scale.} 
   \label{fig:speed}
\end{figure}

We analyze experimental streaming velocities using (the adaptation by \citet{blair2008matlab} of) the particle tracking algorithms of \citet{crocker1996methods}. From the experimental videos, we  calculate the speeds of microparticles, which we approximate by the magnitude of the horizontal velocity as the vertical velocity is smaller. The results of our analysis are shown in figure \ref{fig:speed}. We first identify particles in a narrow sector, concentric with the cylinder, of angular width $2\delta \theta$ about a mean angle $\theta$ (we pick $\theta = \pi/4$, $\delta\theta=0.1$). We subdivide this sector into multiple radial sections (typically 30) starting from the cylinder surface. We collect hundreds of particle trajectories in the sector of interest and calculate their positions and speeds as they pass through the angle $\theta$ (light gray markers in figure \ref{fig:speed}). We then average the particle positions and speeds within each radial section of the sector, which we identify with the mean speed at the projected averaged locations of the particles $v(r, \theta)$ (colored markers in figure \ref{fig:speed}). Deviations from the average thus represent the spread of the individual particle positions and speeds. The maximum (bin-averaged) speed for each experiment is denoted $v_{\rm m}$ and its radial location is denoted $r_{\rm m}$. 

In figure \ref{fig:speed}, we plot the ratio of (bin-averaged) speed in each segment and maximal speed among all particles, as a function of the dimensionless distance to the cylinder axis. The experimental data are collected from three characteristic experiments with cylinders confined axially in microchannels: (i) with height $2h=480\,\mu m$ and cylinder radius $a=600\,\mu m$, operating at a frequency of $f = 100$ Hz ($\eta=0.4$, $\delta = 0.23$; see left panel of figure \ref{fig:StreamlinePlot}(a)); (ii)  $2h=480\,\mu m$, $a=750\,\mu m$, driven at $f = 50$ Hz ($\eta=0.32$, $\delta = 0.33$); (iii) $2h=160\,\mu m$, $a=600\,\mu m$, operating at $f = 300$ Hz ($\eta=0.13$, $\delta = 0.4$; see right panel of figure \ref{fig:StreamlinePlot}(a) and figure \ref{fig:StreamlinePlot}(b)). As shown in Figure \ref{fig:experiment} and discussed earlier, the large depth of field picks up particles at different depths, encompassing a wide range of speeds (including speeds close to zero at depths where flow reverses); our data set includes all of these trajectories. 

Figure \ref{fig:speed}(a) shows the experimental data for $v/v_{\rm m}$ versus $r$.  As our theory predicts, the streaming velocity is found to  decay as the inverse cube of the distance from the cylinder ($|\bm{v}_2|\propto r^{-3}$). 
Parameters $\epsilon$, $\eta$ and $\delta$ are different from experiment to experiment, resulting in different $v_{\rm m}$ and $r_{\rm m}$. We observe that the theory predicts $r_{\rm m}=1$ at $\eta\rightarrow 0$, which is consistent with the experiment for smallest $\eta$. 

Although the theory is formally only accurate for small $\eta$ we find that it nonetheless provides a useful way to organize all of the data onto a universal curve; figure \ref{fig:speed}(b). We assume that the  maximum binned mean speed from experimental data exists in the  outer region, even for moderate $\eta$. The theory predicts a velocity decay behavior $\emph{v}\propto r^{-3}$ (with a prefactor that generally depends on $\delta$ and $\eta$), and so we anticipate that the maximal velocity satisfies $v_{\rm m}\propto r^{-3}_{\rm m}$, with the same prefactor.  Rearranging the equation yields a universal curve $v/(v_{\rm m} r^3_{\rm m}) = 1/r^3$. Applying this relationship, we find that the rescaled experimental data in figure \ref{fig:speed}(b) closely follows this universal curve across two orders of magnitude, though the agreement is better with an additional prefactor of $1.1$. 

\section{Discussion and conclusions}

In this paper, we have developed a three-dimensional streaming theory around cylindrical obstacles confined axially in Hele-Shaw-like microhannels. We exploited the separation between vertical and horizontal lengths scales and utilized lubrication theory under the conditions of small oscillation amplitudes but retaining the inertia of the flow. After resolving the primary (oscillatory) and secondary (steady) flow, we have investigated the 3D streaming structure around the cylinder which has two unique characteristics compared with  previous research: ($\rm i$) The flow direction varies across the depth of the channel, and ($\rm ii$) The streaming velocity decays with the inverse cube of the distance from the cylinder center. These findings were validated experimentally driving oscillations around cylinders of different radii sandwiched inside PDMS microchannels of various aspect ratios. We verified the vertical reversal of horizontal flow by analyzing the trajectories of suspended tracer microparticles and the decay of flow velocity by particle tracking velocimetry measurements.

As noted in section \ref{section3}, the theory developed here focuses on flow in an ``outer region'' wherein distances from the cylinder surface are much greater than the channel height. This is only part of the picture, as a separate ``inner region'' surrounding the cylinder surface (and with radial width comparable to the channel height) must also exist for the flow to correctly satisfy the no-slip conditions at the cylinder surface. Within the inner region where $r-1=O(\eta)$, radial variations of the flow are as important as axial ones. In appendix A, we analyze the variation of velocity within this inner region, which in turn produce a small $O(\eta)$ normal velocity which must be met by the outer flow as a matching condition at $r = 1$.  We anticipate that the correction to the secondary flow is of similar size and can be similarly computed though the details are significantly more complex and are left to future work. We find (Appendix A) that one important qualitative feature of accounting for the inner region is the introduction of steady flow components of $O(\eta)$ with nonzero channel-averaged velocity (recall that the steady flow at $O(\eta^0)$ has zero channel average). Such a channel-averaged flow is likely be important for transport, in particular when $h/a$ is moderately large, but is beyond the focus of this work. 

Previous work on three-dimensional streaming focused on relatively deep channels \citep{marin2015three,volk2020size,rallabandi2015three} and report 
systematic small axial displacements following quasi-planar orbits on a toroidal surface. 
By contrast, the streaming that we have studied in relatively shallow channels exhibits a qualitatively different structure involving an additional axial re-circulation that is necessary to maintain the small depth-averaged flow predicted by the theory. 
The ability to generate such recirculating three-dimensional flows in microconfined geometries is promising for particle manipulation and micromixing applications.  For a much wider range of microstreaming set-ups, the tools developed here will be helpful to assess, model, and either minimize or enhance three-dimensional flow effects 
in practical applications.

\subsection*{Acknowledgements}
We thank S. Hilgenfeldt, A. Gupta,  W. Grover, M. Rao, J. Sheng and H. Tsutsui for stimulating discussions. X. Z. acknowledges partial support of Dean's Distinguished Fellowship from the UCR Graduate Division. The authors gratefully acknowledge support from the National Science Foundation under grant number CBET 2143943. \\

\noindent \textbf{Declaration of Interests.} The authors report no conflict of interest.
\appendix
\section{Inner region}\label{appA}
We resolve some details of the inner region where $r - 1 = O(\eta)$, which also determines the effective boundary conditions for the outer flow at $r=1$.  We define a rescaled inner coordinate $s=(r-1)/\eta$ that is $O(1)$ in the inner region. The continuity equation written in cylindrical coordinates, is $r^{-1} \partial \left(r v_r^{\rm tot}\right)/{\partial r} +r^{-1}\partial v_\theta^{\rm tot}/{\partial \theta} +\partial v_z^{\rm tot}/{\partial z} =0$, where the superscript ``tot" refers to the total or exact solution to the flow comprising both outer and inner contributions (for consistency with the main text, outer flow quantities have no superscripts). Integrating the continuity equation across the inner region and through the channel depth (and changing variables from $r$ to $s$), we obtain the exact relation
\begin{align} \label{ContIntegrate}
    r \overline{v^{\rm tot}_r} \Big|_{s = 0}^{\infty} = -\eta \int_0^{\infty} \frac{\partial \overline{v^{\rm tot}_\theta}}{\partial \theta} \rmd s.
\end{align}
An analogous result can also be obtained for the outer solution. Combining \eqref{ContIntegrate} with its outer analog, noting that the $\bm{v}^{\rm tot}$ vanishes at the cylinder surface ($s = 0$) and that $\bm{v}^{\rm tot} = \bm{v}$ at $s = \infty$, we obtain the effective condition for the outer flow at the surface of the cylinder,
\begin{align} \label{NormalVelBCInner}
    \overline{\bm{v}}\cdot\bm{n} = -\frac{\eta}{r} \frac{\partial}{\partial \theta}\int^\infty_0\overline{v_{\theta}^{\rm tot} - v_{\theta}} \, \rmd s\quad\mbox{at}\quad r=1,
\end{align}
which reduces to (\ref{nondimNSbc}c) at leading order for $\eta \ll 1$. The right hand side represents the flux from the inner region that leaks into the outer flow and produces an $O(\eta)$ correction to (\ref{nondimNSbc}c). 
\begin{figure}
\centering
\includegraphics[width=0.8\textwidth]{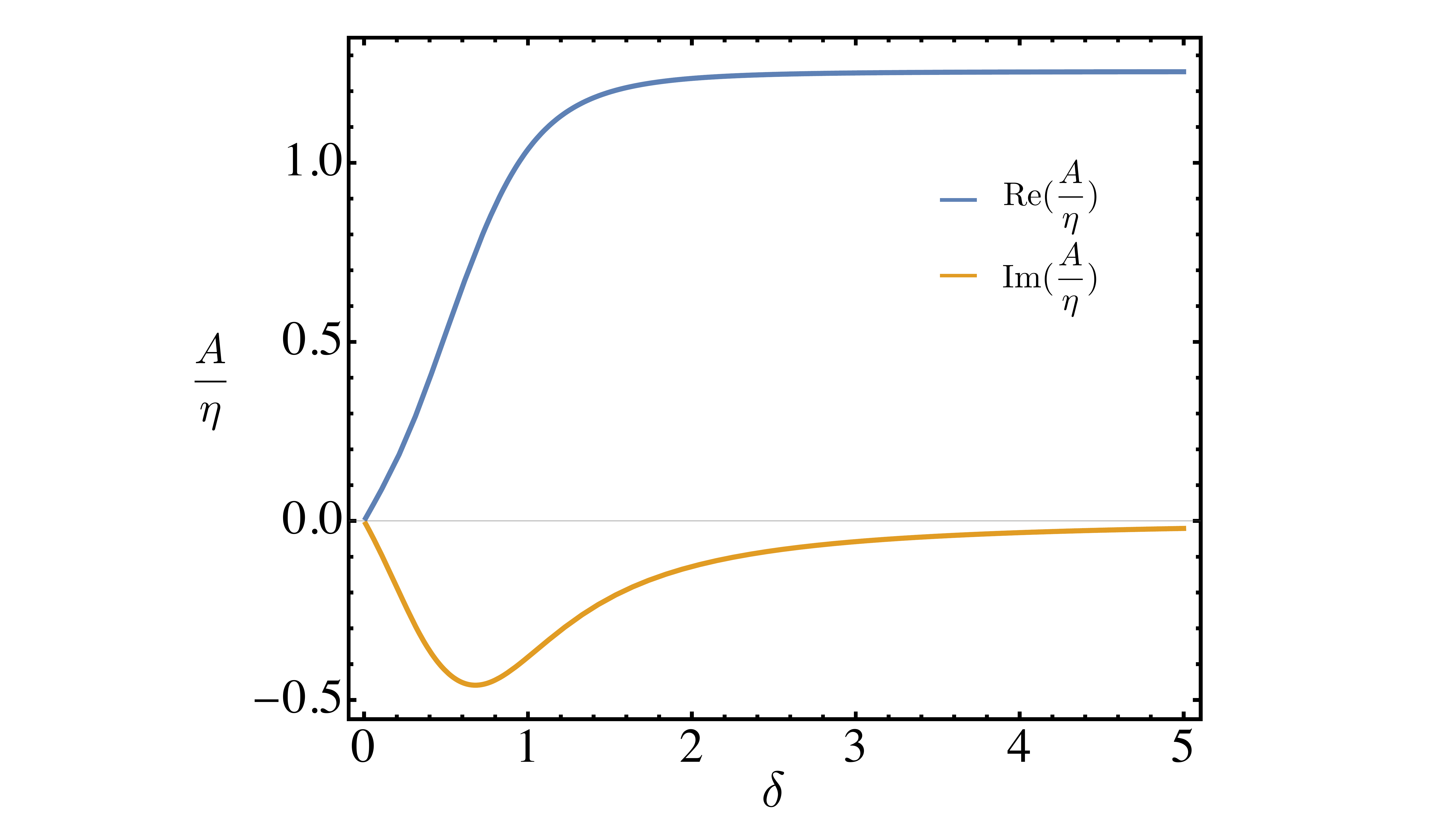}
\caption{\label{fig:Innercorrection}Primary pressure correction $A$ versus the Stokes layer thickness $\delta$.}
\label{approx}
\end{figure}

\subsection{Primary flow}
We calculate these $O(\eta)$ corrections for the primary flow and then discuss some implications on the streaming. The primary flow in the outer region (where $r-1\gg \eta$) is governed by \eqref{prifloweq}, except for the boundary condition at $r=1$, which will be refined using \eqref{NormalVelBCInner}. Ignoring the boundary condition at $r=1$, the primary outer pressure is 
\begin{align}\label{innerpressure}
    p_1 = G(\alpha) \left(r+\frac{(1+A)}{r}\right) \cos{\theta}\, \mathrm{e^{i t}},
\end{align}
and the velocities are related to pressure in the same way as in \eqref{priflowsolu}. 

To find $A$ up to $O(\eta)$, we first calculate the inner solution for the primary azimuthal velocity. Using standard analysis techniques in the inner region (where $s = (r -1)/\eta = O(1)$), we find that the pressure  remains independent of $z$ and does not vary significantly across the inner region. Then, the primary azimuthal velocity $v_{\theta 1}^{\rm tot}$ (accounting for both inner and outer contributions) is governed by 
\begin{align} \label{UthetaInnerGE}
    \alpha^2 v_{\theta 1}^{\rm tot} =-\frac{1}{r}\frac{\partial p_1}{\partial \theta}+\left(\frac{\partial^2}{\partial s^2}+\frac{\partial^2}{\partial z^2}\right) v^{\rm tot}_{\theta 1}.
\end{align}
Solving \eqref{UthetaInnerGE} subject to no-slip conditions at the walls at $z = \pm 1$ yields
\begin{align}\label{uthe0total}
    v_{\theta 1}^{\rm tot} = \left[-\frac{1}{\alpha^2}\left(1-\frac{\cosh{\alpha z}}{\cosh{\alpha}}\right)+\sum^\infty_{n=1} c_n\cos{k_n z}\;\mathrm{e}^{-s \sqrt{k_n^2+\alpha^2}}\right]\frac{1}{r}\frac{\partial p_1}{\partial \theta},
\end{align}
where $k_n=(2n-1)\pi/2$ and the coefficients $c_n$ are to be determined.  Observe that  $v_\theta^{\rm tot}$ comprises the outer solution (first term) plus an inner correction that decays exponentially with $s$ (second term). The no-slip condition at $r=1\, (s=0)$ yields the coefficients $c_n=-2 \cos (\pi  n)/(k_n (k_n^2 + \alpha^2)).$ Substituting $c_n$ and \eqref{uthe0total} back into \eqref{NormalVelBCInner}, we obtain the condition $\overline{\bm{v}}_{1}\cdot\bm{n} = -\eta (\mathcal{L}/2)\, \partial^2 p_1/\partial \theta^2$ at $r = 1$, where  $\mathcal{L}(\alpha) =\sum^\infty_{n=1}4 k_n^{-2}(k_n^2 + \alpha^2)^{-3/2}$. Comparing with the outer solution ($\overline{\bm{v}}_{1} = \nabla p_1/G$), we find 
\begin{align}
    A=\frac{\eta\alpha^3\mathcal{L}(a)}{\alpha-\tanh{(\alpha)}},
\end{align}
which provides the leading correction to the pressure \eqref{innerpressure} due to the inner flow. The relationship between the correction $A$ and $\delta$ is shown in figure 6.

\subsection{Stokes drift}
Using  (\ref{stokedrift}) in the main text and applying the revised primary  pressure (\ref{innerpressure}), we obtain the Stokes drift in the outer region as  
\begin{align}
    \bm{v}_d= (v_{d r}, v_{d \theta}, \eta v_{d z}) = - \mbox{Im}(A)|G|^2\left|\frac{1}{\alpha^2}\left(1-\frac{\cosh{\alpha z}}{\cosh{\alpha}}\right)\right|^2\left(\frac{\cos{2 \theta}}{r^3}, \frac{\sin{2 \theta}}{r^3}, 0\right),
\end{align}
The Stokes drift is $O(\eta)$ and is thus weaker than the streaming of the main text, but has nonzero average across the channel. Furthermore, the Stokes drift also decays as the inverse cube of the distance from the cylinder, which is the same decay rate found of the leading-order streaming flow. 

\bibliographystyle{jfm}
\bibliography{jfm-instructions}

\end{document}